\documentclass[a4paper,aps,prd,onecolumn,preprintnumbers,showpacs,nofootinbib]{revtex4}
\usepackage{amsmath,amssymb,graphics,epsfig,subfigure}
\usepackage{color}

\begin{document}
\newcommand {\nn} {\nonumber}
\renewcommand{\baselinestretch}{1.3}

\title{Null geodesics and gravitational lensing in a nonsingular spacetime}
\author{Shao-Wen Wei$^{1}$\footnote{weishw@lzu.edu.cn},
        Yu-Xiao Liu$^{1}$\footnote{liuyx@lzu.edu.cn}, and
        Chun-E Fu$^{2}$\footnote{fuche13@mail.xjtu.edu.cn}}

\affiliation{$^{1}$ Institute of Theoretical Physics, Lanzhou University, Lanzhou 730000, People's Republic of China\\
$^{2}$ School of Science, Xi'an Jiaotong University, Xi'an 710049, People's Republic of China}

\begin{abstract}
In this paper, the null geodesics and gravitational lensing in a nonsingular spacetime are investigated. According to the nature of the null geodesics, the spacetime is divided into several cases. In the weak deflection limit, we find the influence of the nonsingularity parameter $q$ on the positions and magnifications of the images is negligible. In the strong deflection limit, the coefficients and observables for the gravitational lensing in a nonsingular black hole background and a weakly nonsingular spacetime are obtained. Comparing these results, we find that, in a weakly nonsingular spacetime, the relativistic images have smaller angular position and relative magnification, but larger angular separation than that of a nonsingular black hole. These results might offer a way to probe the spacetime nonsingularity parameter and put a bound on it by the astronomical instruments in the near future.
\end{abstract}

% \Keywords{ }


\maketitle

\section{Introduction}

The cosmic censorship hypothesis \cite{Penrose,Hawking1} says that singularities that arise in the solutions of Einstein's equations are typically hidden within event horizons, and therefore cannot be seen from the rest of spacetime. However, in a semiclassical approximation \cite{Hawking}, black holes tend to shrink until the central singularities are reached, which will lead to the breakdown of the theory.

Motivated by the idea of the free singularities, there are several ways to obtain black hole spacetime with no singularities at the center. The one presented in \cite{Nicolini2005plb,Nicolini2006plb} was inspired by the noncommutative geometry. The points on the classical commutative manifold are replaced by states on a noncommutative algebra, and the point-like objects are replaced by smeared objects. Thus the singularity problem is cured at the terminal stage of black hole evaporation.

Another way is to introduce a de Sitter core to replace the central singularity. The first one constructed in this way is the Bardeen regular black hole \cite{Bardeen,Borde94,Borde97,Ayon-Beato00}, which was found to have both an event horizon and a Cauchy horizon. Recently, Hayward proposed a nonsingular black hole solution \cite{Hayward} (Poisson and Israel also derived an equivalent solution based on a simple relation between vacuum energy density and curvature \cite{Poisson,Poisson2}), which is a minimal model satisfying the asymptotically flat and flatness conditions at the center. The static region of it is Bardeen-like. In this nonsingular spacetime, a black hole could generate from an initial vacuum region and then subsequently evaporate to a vacuum region without singularity \cite{Hayward}. This case was extended to the $d$+1 dimensional spacetime and some interesting results were obtained \cite{Spallucci}. The quasinormal frequency of this nonsingular spacetime has been recently analyzed in \cite{Flachi} with a significant difference from the singular spacetime. In fact, according to the nature of this spacetime, we can divide it into several cases, i.e., the nonsingular black hole, the extremal nonsingular black hole, the weakly nonsingular spacetime, the marginally nonsingular spacetime, and the strongly nonsingular spacetime. Other regular black hole solutions \cite{Cataldo,Burinskii,Bronnikov,Berej,Dymnikova,Beato,Lemos,Ansoldi} can be constructed with the introduction of some external form of matter, such as nonlinear magnetic monopole, electrodynamics, or gaussian sources, which leads to that they are not vacuum solutions of Einstein's equations.

The subject of gravitational lensing by black holes and compact stars has received great attention in the last ten years, basically due to the strong evidence that the presence of supermassive black holes at the center of galaxies. The study can be traced back to \cite{Darwin}, where the author examined the gravitational lensing when the light passing near the photon sphere of Schwarzschild spacetime. In Ref. \cite{Narasimha}, the authors showed that, in the case of large values of the scalar charge, the lensing characteristics were significantly different. And the result provides a preliminary knowledge on the naked singularity lens. This resurrects the study of the gravitational lensing. After modeled the massive dark object at
the Galactic center as a Schwarzschild black hole lens, it was found that \cite{Virbhadra,Virbhadra2,Virbhadra2009}, similar to the Darwin's paper, apart from a primary image and a secondary image resulting by small bending of light in a weak gravitational field, there is theoretically an infinite sequence of very demagnified images on both sides of the optical axis. Similar result was also found in \cite{EiroaRomero,Perlick}. These images were named as the ``relativistic images" by Virbhadra and Ellis \cite{Virbhadra} and that term was extensively used in later work. Based on lens equation \cite{Frittelli,Virbhadra}, Bozza \emph{et al}. \cite{Bozza,Bozza03prd,Bozza01,Bozza02,Bozza03} developed a semi-analytical method to deal with it. This method has been applied to other black holes \cite{BozzaBozza,Eiroa,Sarkar06,ChenJing09,Eiroa66,Eiroa12,Eiroa13,
Whisker,Bhadra,Vazquez,Nandi,Gyulchev75,Gyulchev13,Bozza76,Virbhadra77,
Bisnovatyi,Nun,LiuChenJing,Ding2011,Dingjing,Liu2012,Ghosh,WeiLiu,WeiLiu2012,Sahu,Perlick2}. These results suggest that, through measuring the relativistic images, gravitational lensing could act as a probe to these black holes, as well as a profound verification of alternative theories of gravity in the strong field regime \cite{Eiroa,Sarkar06,ChenJing09}. Furthermore, it can also guide us to detect the gravitational waves at proper frequency \cite{Stefanov,Weiprdr}. It is also worthwhile to mention that in Ref. \cite{Virbhadra2009}, the author pointed out that Bozza's semi-analytical method gives small percentage difference of the deflection angle, angular position, and angular separation than their accurate values, while large percentage difference of magnification and differential time delays among the relativistic images. Thus one must pay a great attention on studying the differential time delays among the relativistic images, and we will not consider that case in this paper.

In Refs. \cite{Ding2011,Dingjing}, the authors studied the strong gravitational lensing by a regular black hole with noncommutative corrected. The result shown that gravitational lensing in the strong deflection limit could provide a probe to the noncommutative parameter. In this paper, we mainly focus on the exploration of the lensing features in a nonsingular spacetime with the central singularity replaced by a de Sitter core. At first, we study the nature of the spacetime in different range of the nonsingularity parameter $q$. And according to it, the spacetime is classified into the nonsingular black hole $q/2M\in (0,\; Q_{\text{cr1}})$, the extremal nonsingular black hole $q/2M=Q_{\text{cr1}}$, the weakly nonsingular spacetime $q/2M\in (Q_{\text{cr1}},\;Q_{\text{cr2}})$, the marginally nonsingular spacetime $q/2M=Q_{\text{cr2}}$, and the strongly nonsingular spacetime $q/2M\in (Q_{\text{cr2}},\;\infty)$. Then under this classification, we study the lensing features in a nonsingular spacetime both in weak and strong deflection limits. The result shows that in the weak deflection limit, the influence of the nonsingularity parameter $q$ on the lensing is negligible. Comparing with it, $q$ has a significant effect in the strong deflection limit, which are very helpful for detecting the nonsingularity of our universe in the future astronomical observations.

The paper is structured as follows. In Sec. \ref{geodesic}, we study the null geodesics and photon sphere for this nonsingular spacetime. In Sec. \ref{deflection}, the influence of the nonsingularity parameter $q$ on the lensing in the weak and strong deflection limits are investigated, respectively. In Sec. \ref{observational}, supposing that the gravitational field of the supermassive black hole at the center of our Milky Way can be described by the nonsingular metric, we estimate the numerical values of the coefficients and observables for gravitational lensing in the strong deflection limit. A brief discussion is given in Sec. \ref{Summary}.

\section{Null geodesics and photon sphere}
\label{geodesic}

In Ref. \cite{Hayward}, Hayward suggested that a nonsingular spacetime, as a minimal model, can be described by the metric
\begin{eqnarray}
 ds^{2}=-f(r)dt^{2}+f(r)^{-1}dr^{2}+r^{2}(d\theta^{2}+\sin^{2}\theta d\phi^{2}),
 \label{metric}
\end{eqnarray}
where the metric function $f(r)$ reads
\begin{eqnarray}
 f(r)=1-\frac{2M r^{2}}{r^{3}+2q^{2}M},\label{metricfunction}
\end{eqnarray}
and its behavior is
\begin{eqnarray}
 f(r)&\sim & 1-\frac{2M}{r}+\frac{4q^{2}M^{2}}{r^{4}}+\mathcal{O}(\frac{1}{r})^{7}
 \quad\mbox{as}\; r\rightarrow \infty,\\
 f(r)&\sim & 1-\frac{r^{2}}{q^{2}}+\frac{r^{5}}{2q^{4}M}+\mathcal{O}(r)^{7}
 \quad\;\;\mbox{as}\; r\rightarrow 0.
\end{eqnarray}
It is quite clear that the spacetime described by the above metric is similar to a Schwarzschild spacetime at large distance. While at small distance, there is an effective cosmological constant, which leads to regularity at $r=0$. The parameter $q$ is a new fundamental constant on the same ground as $\hbar$ and $c$. In order to keep some degree of generality, we consider $q$ as a free, model-dependent parameter. On the other hand, it is clear that, when $q=0$, there will be singular at $r=0$. So, we can name $q$ as a nonsingularity parameter measuring the nonsingularity of a spacetime.

The outer and inner horizons $r_{\pm}$ are determined by $f(r)=0$. And for the spacetime with large mass, we approximately have $r_{+}=2M$ and $r_{-}=q$. In order to compute the null geodesics in this nonsingular spacetime, we follow \cite{Chandrasekhar}. Here, we only restrict our attention to the equatorial orbits with $\theta=\pi/2$. The Lagrangian is
\begin{eqnarray}
 2\mathcal{L}=-f(r)\dot{t}^{2}+\dot{r}^{2}/f(r)+r^{2}\dot{\phi}^{2}.
 \label{lagrangian}
\end{eqnarray}
The generalized momentum can be defined from this Lagrangian as $p_{\mu}=\frac{\partial \mathcal{L}}{\partial \dot{x}^{\mu}}=g_{\mu\nu}\dot{x}^{\nu}$ with its components given by
\begin{eqnarray}
 p_{t}   &=&-f(r)\dot{t}\equiv -E=\text{const},\label{pt}\\
 p_{\phi}&=&r^{2}\dot{\phi}\equiv L=\text{const},\label{phi}\\
 p_{r}   &=&\dot{r}/f(r).
\end{eqnarray}
Substituting Eqs. (\ref{pt}) and (\ref{phi}) into (\ref{lagrangian}), we find that the Lagrangian $\mathcal{L}$ is independent of both $t$ and $\phi$. Thus, we immediately get two integrals of the motion $p_{t}$ and $p_{\phi}$. Solving (\ref{pt}) and (\ref{phi}), we easily obtain the $t$ motion and $\phi$ motion,
\begin{eqnarray}
 \dot{t}&=&\frac{E}{f(r)},\\
 \dot{\phi}&=&\frac{L}{r^{2}\sin^{2}\theta}.\label{phit2}
\end{eqnarray}
The Hamiltonian is given by
\begin{eqnarray}
 2\mathcal{H}=2(p_{\mu}\dot{x}^{\mu}-\mathcal{L})
     &=&-f(r)\dot{t}^{2}+\dot{r}^{2}/f(r)+r^{2}\dot{\phi}^{2}\nonumber\\
     &=&-E\dot{t}+L\dot{\phi}+\dot{r}^{2}/f(r)=\delta.
\end{eqnarray}
Here $\delta$ is another integral of the motion. And $\delta=-1,0,1$ are for spacelike, null and timelike geodesics, respectively. Since we consider the null geodesics, so we choose $\delta=0$ here. Then the radial motion can be expressed as
\begin{eqnarray}
 \dot{r}^{2}+\mathcal{V}_{\text{eff}}(r)=E^{2},\label{rt2}
\end{eqnarray}
with the effective potential $\mathcal{V}_{\text{eff}}=\frac{L^{2}}{r^{2}}(1-\frac{2Mr^{2}}{r^{3}+2q^{2}M})$. Then the circular geodesics satisfy
\begin{eqnarray}
 \mathcal{V}_{\text{eff}}(r)=E^{2},\quad \frac{\partial \mathcal{V}_{\text{eff}}}{\partial r}=0.
 \label{veff}
\end{eqnarray}
Moreover, a stable (unstable) circular orbit requires $\frac{\partial^{2}\mathcal{V}_{\text{eff}}}{\partial r^{2}}>0$ ($<$0), which admits a minimum (maximum) of the effective potential. Solving Eq. (\ref{veff}), we have
\begin{eqnarray}
 2f(r)-rf'(r)=0. \label{photon1}
\end{eqnarray}
And for the metric (\ref{metricfunction}), this equation reduces to
\begin{eqnarray}
 2r^{6}-3r^{5}+4q^{2}r^{3}+2q^{4}=0,\label{photon11}
\end{eqnarray}
where, for simplicity, we measure all quantities with the Schwarzschild radius, which is equivalent to putting $2M = 1$ in all equations.
\begin{figure}
\centerline{\includegraphics[width=8cm]{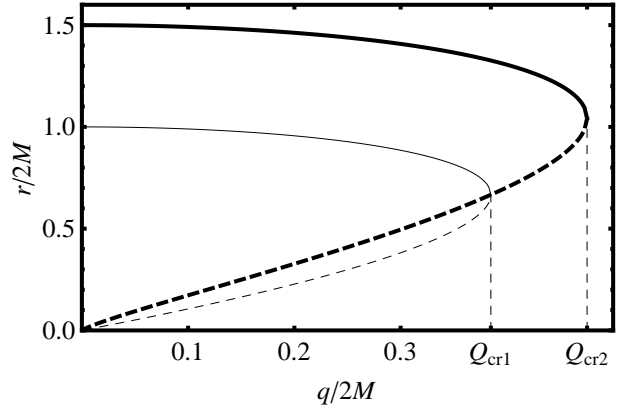}}
\caption{Horizons and circular orbits for the nonsingular spacetime. The thin solid and dashed lines are for the outer and inner horizons, respectively. And the thick solid and dashed lines are for the photon sphere and stable circular orbit, respectively. The parameters $Q_{\text{cr1}}=(2\sqrt{3})/9=0.385$ and $Q_{\text{cr2}}=(25\sqrt{30})/288=0.475$.} \label{Rps}
\end{figure}
Solving (\ref{photon11}), we can obtain the stable and unstable circular orbits for this nonsingular spacetime. For a spherically symmetric and static spacetime, the photon sphere is known as an unstable circular orbit of photon (other definitions can be found in \cite{Virbhadra,Claudel}). So, we can obtain the photon sphere for this nonsingular spacetime from (\ref{photon11}) with the unstable condition $\frac{\partial^{2}\mathcal{V}_{\text{eff}}}{\partial r^{2}}<0$. It is obvious that this relation (\ref{photon11}) is quite different from that in the Schwarzschild black hole spacetime, which implies that, in the strong field limit, there exist some distinct effects of $q$ on the gravitational lensing. The stable circular orbit can also be got by imposing $\frac{\partial^{2}\mathcal{V}_{\text{eff}}}{\partial r^{2}}>0$. The event horizons, photon sphere, and stable circular orbit are plotted in Fig. \ref{Rps} as a function of $q/2M$. It is clear that there are several distinct ranges of the parameter emerging where the structures of the horizons and circular geodesics will be qualitatively different, namely $q/2M\in (0,\; Q_{\text{cr1}})$, $q/2M=Q_{\text{cr1}}$, $q/2M\in (Q_{\text{cr1}},\;Q_{\text{cr2}})$, $q/2M=Q_{\text{cr2}}$, and $q/2M\in (Q_{\text{cr2}},\;\infty)$. In the following we will discuss these cases, respectively.\\

%%%%%%%%%%%%%%%%%%%%%%%%%%%%%%%%%%%%%%%%%%%%%%%%%%%%%%%%%%%%%%%%%%%%%%%%%%%%%
\begin{figure*}
\subfigure[the nonsingular black hole]{\label{V1}
\includegraphics[width=8cm]{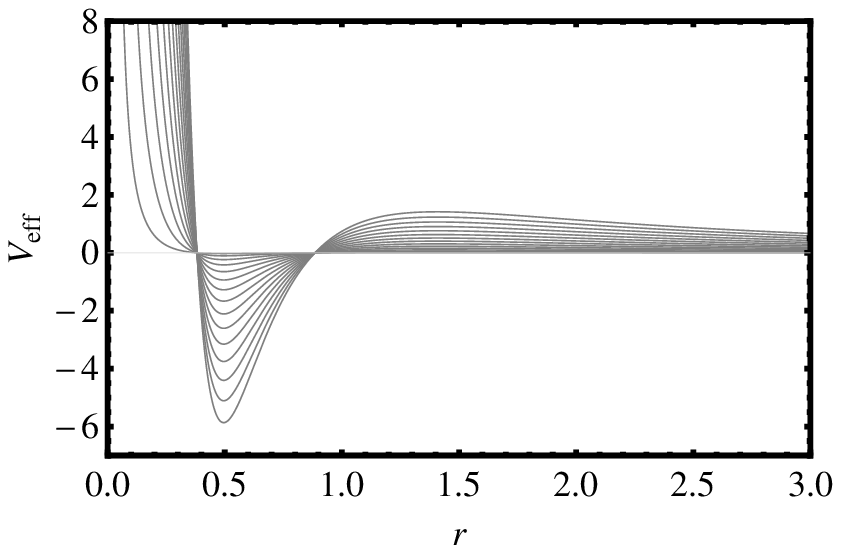}}
\subfigure[the extremal nonsingular black hole]{\label{V2}
\includegraphics[width=8cm]{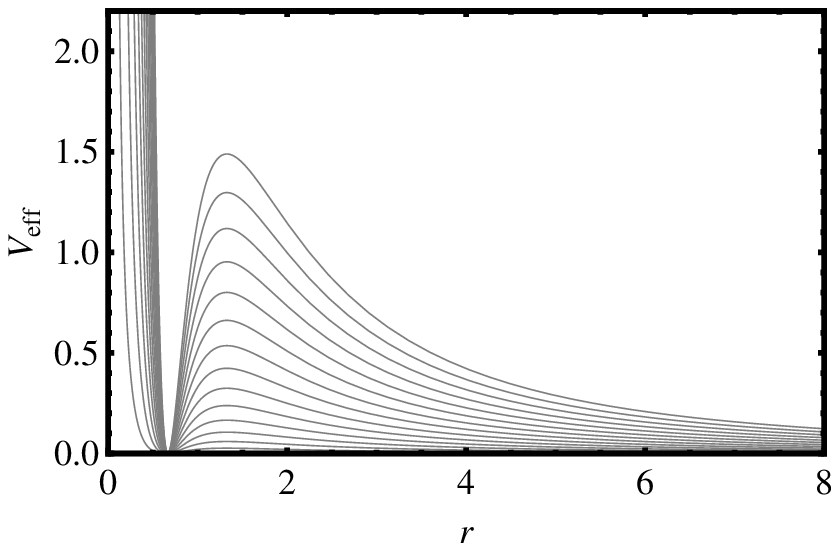}}\\
\subfigure[the weakly nonsingular spacetime]{\label{V3}
\includegraphics[width=8cm]{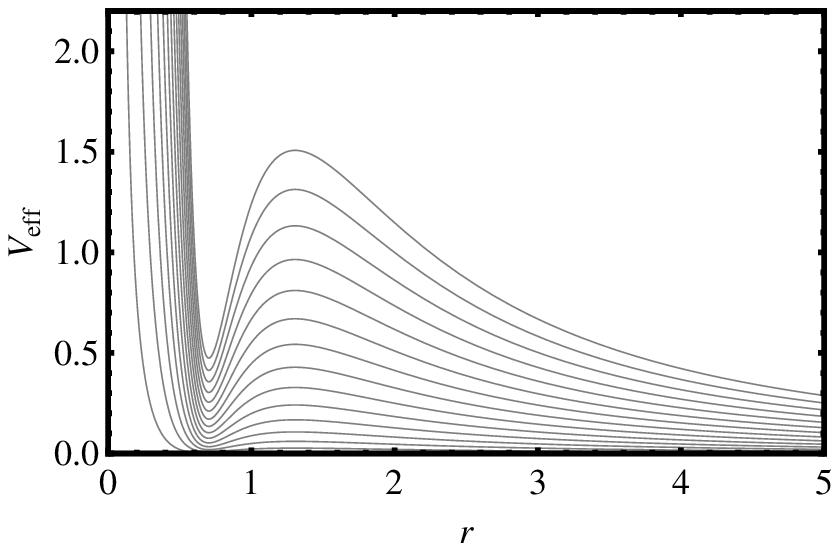}}
\subfigure[the marginally nonsingular spacetime]{\label{V4}
\includegraphics[width=8cm]{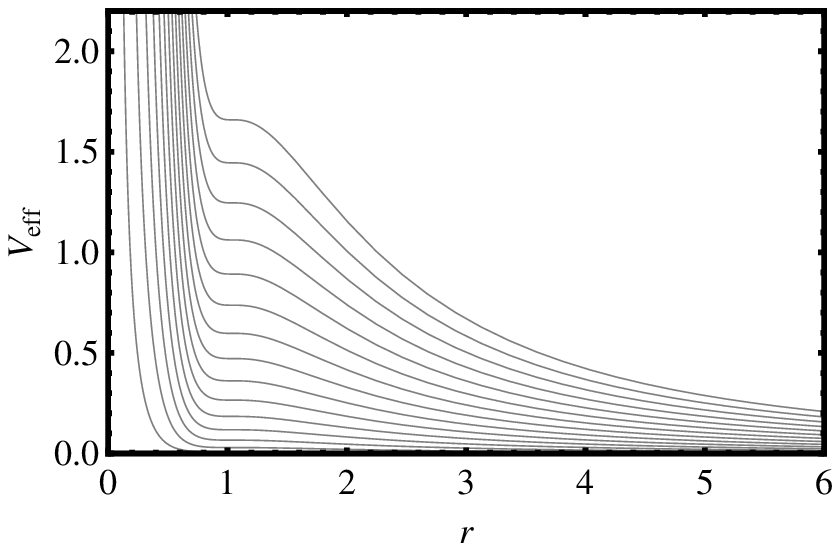}}\\
\centerline{\subfigure[the strongly nonsingular spacetime]{\label{V5}
\includegraphics[width=8cm]{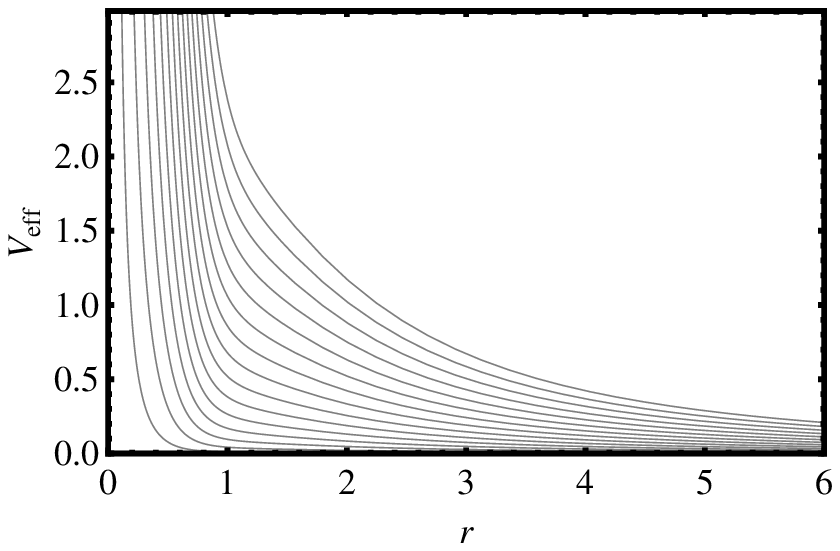}}}
\caption{Behavior of the effective potential $\mathcal{V}_{\text{eff}}$ as a function of $r$ with different values of $L$.} \label{V}
\end{figure*}
%%%%%%%%%%%%%%%%%%%%%%%%%%%%%%%%%%%%%%%%%%%%%%%%%%%%%%%%%%%%%%%%%%%%%%%%%%%%%%%%

\noindent\textbf{Case 1}: nonsingular black hole $q/2M\in (0,\; Q_{\text{cr1}})$

Since $\mathcal{V}_{\text{eff}} \sim f(r)$, the horizon locates at $\mathcal{V}_{\text{eff}}=0$. And note that the local minimum and maximum of $\mathcal{V}_{\text{eff}}$ correspond to the stable and unstable circular orbits, respectively. Then we can easily read the property of different cases from Fig. \ref{V}.

For the case of a nonsingular black hole with $q/2M\in (0,\; Q_{\text{cr1}})$, there are two horizons $r_{\pm}$, one photon sphere $r_{\text{ps}}$, and one stable circular orbit $r_{\text{cir}}$. And they satisfy the following relation
\begin{eqnarray}
 r_{-}<r_{\text{cir}}<r_{+}<r_{\text{ps}}.
\end{eqnarray}
This implies that the outer horizon is always covered by a photon sphere, and the stable circular orbit locates between these two horizons.

In Fig. \ref{V1}, the general behavior of the effective potential $\mathcal{V}_{\text{eff}}(r)$ is shown as a function of $r$ for different values of the angular momentum $L$. We find that the effective potential admits two zeros corresponding to the outer and inner horizons, as well as one maximum and one minimum corresponding to the photon sphere and stable circular orbit. We can also see that the local minimum point always lies between these two zeros indicating the stable circular orbit lies in the region between the two horizons.
\\

\noindent\textbf{Case 2}: extremal nonsingular black hole $q/2M=Q_{\text{cr1}}$

For this case, we have the following relation
\begin{eqnarray}
 r_{-}=r_{+}=r_{\text{cir}}<r_{\text{ps}}.
\end{eqnarray}
The first `$=$' means that the two horizons coincide with each other. This case corresponds to an extremal nonsingular black hole. The second `$=$' implies that the degenerate horizon is also a stable circular orbit against small perturbation.

The detailed behavior of the effective potential $\mathcal{V}_{\text{eff}}(r)$ is presented in Fig. \ref{V2}. It shows that the effective potential have one zero and one minimum located at the same point. It also admits a maximum corresponding to the photon sphere.
\\

\noindent\textbf{Case 3}: weakly nonsingular spacetime $q/2M\in (Q_{\text{cr1}},\;Q_{\text{cr2}})$

For the case of $q/2M\in (Q_{\text{cr1}},\;Q_{\text{cr2}})$, we clearly see from Fig. \ref{Rps} that the horizon disappears. So this case describes a nonsingular spacetime without a black hole. Obviously, the radius of the photon sphere and stable circular orbit satisfy
\begin{eqnarray}
 r_{\text{cir}}<r_{\text{ps}}.
\end{eqnarray}
This tells that the stable circular orbit is always covered by a photon sphere. The effective potential $\mathcal{V}_{\text{eff}}(r)$ is plotted in Fig. \ref{V3}, from which we find that $\mathcal{V}_{\text{eff}}(r)$ is positive for all $r$. Thus, the horizons do not exist in this case, which is consistent with the result from Fig. \ref{Rps}. On the other hand, there are one minimum and one maximum corresponding to the stable circular orbit and photon sphere, respectively.
\\

\noindent\textbf{Case 4}: marginally nonsingular spacetime $q/2M=Q_{\text{cr2}}$

For this case, the horizon also disappears. And the stable and unstable circular orbits coincide with each other, i.e., $r_{\text{cir}}=r_{\text{ps}}$. For this circular orbit, we have $\frac{\partial^{2}\mathcal{V}_{\text{eff}}}{\partial r^{2}}=0$. This result can be seen from Fig. \ref{V4}. One the other hand, we will see in next section that some strong deflection limit coefficients will diverge in this case caused by $r_{\text{cir}}=r_{\text{ps}}$.
\\

\noindent\textbf{Case 5}: strongly nonsingular spacetime $q/2M\in (Q_{\text{cr2}},\;\infty)$

For the last case, all the horizons, photon sphere, and the stable circular orbit disappear. The effective potential $\mathcal{V}_{\text{eff}}(r)$ is plotted in Fig. \ref{V5}. It monotonically decreases from infinity to zero as $r$ goes from 0 to $\infty$.

From the above  discussion, we find that the photon sphere of a nonsingular black hole is always larger than that of a weakly nonsingular spacetime. Therefore, we come to the conclusion that, a photon is more easily captured by a nonsingular black hole. Since the photon sphere disappears for a strongly nonsingular spacetime, we will not focus on the lensing for a strongly nonsingular spacetime.

\section{Lensing in nonsingular spacetime}
\label{deflection}

In this section, we will study the lensing in nonsingular spacetime. The influcenceof nonsingularity parameter $q$ on the position and magnification of the non-relativistic and relativistic images will be analyzed.

\subsection{The deflection angle and lens geometry}

Taking into account the spherical symmetry of this spacetime, we just consider the case that both the observer and the source lie in the equatorial plane ($\theta=\pi/2$), resulting that the whole trajectory of the photon is also restricted in the same plane. Then the deflection angle for the photon coming from infinity and returning to infinity is
\begin{eqnarray}
 \alpha(r_{0})=I(r_{0})-\pi,\label{deflectionangle}
\end{eqnarray}
with the total azimuthal angle given by
\begin{eqnarray}
 I(r_{0})=2\int_{r_{0}}^{\infty}
   \frac{r_{0}}{r}\frac{dr}{\sqrt{r^{2}f(r_{0})-r_{0}^{2}f(r)}},
\end{eqnarray}
where Eqs. (\ref{phit2}) and (\ref{rt2}) are used and $E$ is set to one. In a black hole spacetime, the deflection angle $\alpha(r_{0})$ is a monotonically decreasing function with $r_{0}$. So it is easy to imagine that the light ray can make a complete loop or more than one loop before reaching the observer. As have been pointed out by other authors that, the value of the deflection angle will be unboundedly large when the photon sphere is reached.

Defining $x=r_{0}/r$, the total azimuthal angle can be expressed as
\begin{eqnarray}
 I(r_{0})=\int_{0}^{1} h(x)dx,\label{az}
\end{eqnarray}
with
\begin{eqnarray}
 h(x)=-2\sqrt{\frac{\left(q^2+r_0^3\right) \left(q^2
   x^3+r_0^3\right)}{\left(1-x^2\right) \left(q^2+r_0^3\right)
   \left(q^2 x^3+r_0^3\right)+r_0^5
   \left(x^3-1\right)}}.\label{hx}
\end{eqnarray}

Here we would like to give a brief study on the lens geometry. The lens configuration is supposed that the black hole is situated between the light source and observer, both of them are far from the black hole lens, so that the gravitational fields are very weak and the spacetime there can be described by a flat metric. An important element of the lens geometry is the optical axis, which is defined as the line joining the observer and the lens. The angular positions of the source (S) and the images (I), seen from the observer, are denoted as $\varpi$ and $\theta$. Then the lens equation reads \cite{Virbhadra}:
\begin{eqnarray}
 \tan\varpi=\tan\theta-\frac{D_{\text{LS}}}{D_{\text{OS}}}\big[\tan(\alpha-\theta)
                    +\tan\theta\big].\label{lensgeometry}
\end{eqnarray}
 $D_{\text{OL}}$, $D_{\text{LS}}$ and $D_{\text{OS}}$ are the observer-lens, lens-source, and observer-source distances, respectively. It was pointed out by Bozza \cite{Bozza08} that, these distances are not the true distances between different positions, however, if the source is very close to the optical axis, they are a reasonably good approximation.

\subsection{Non-relativistic images}

Let us first consider the lensing in the weak deflection limit, where photon has a large impact parameter, i.e., $r_{0}\gg1$. Then the function $h(x)$ in Eq. (\ref{hx}) can be performed a Taylor expansion around $1/r_{0}$, which is given by
\begin{eqnarray}
h(x)= \frac{2}{\sqrt{1-x^2}}
   &+&\frac{x^2+x+1}{(x+1) \sqrt{1-x^2}}\frac{1}{r_{0}}
   +\frac{3 \left(x^2+x+1\right)^2}{4(x+1)^2
   \sqrt{1-x^2}}\frac{1}{r_{0}^{2}}
   +\frac{5 \left(x^2+x+1\right)^3}{8 (x+1)^3
   \sqrt{1-x^2}}\frac{1}{r_{0}^{3}}\nonumber\\
   &+&\frac{35 \left(x^2+x+1\right)^4-64 q^2 (x+1)^4
   \left(x^4+x^2+1\right)}{64 (x+1)^4
   \sqrt{1-x^2}}\frac{1}{r_{0}^{4}}
   +\mathcal{O}\left(\frac{1}{r_{0}^{5}}\right).
\end{eqnarray}
Thus, the deflection angle is calculated as
\begin{eqnarray}
 \alpha(r_{0})= A_{1}\frac{1}{r_{0}}
      +A_{2}\frac{1}{r_{0}^{2}}
      +A_{3}\frac{1}{r_{0}^{3}}
      +A_{4}\frac{1}{r_{0}^{4}}+\mathcal{O}\bigg(\frac{1}{r_{0}^{5}}\bigg)
\end{eqnarray}
with the coefficients
\begin{eqnarray}
   A_{1}=2,\quad
   A_{2}=\frac{15\pi-16}{16},\quad
   A_{3}=-\frac{45\pi-244}{48},\quad
   A_{4}=(2.94524 q^2-2.50549).
\end{eqnarray}
Note that the nonsingularity parameter $q$ only affects the deflection angle in the forth order. Since $r_{0}$ has a large value, the influence of $q$ on the non-relativistic images are very weak. Restoring the dimension, the first three coefficients are exactly consistent with that of Ref.~\cite{Keeton}.

For high alignment, and using the lens equation (\ref{lensgeometry}), the image positions and magnifications in the weak deflection limit can be written as a series expansion of the form \cite{Keeton}
\begin{eqnarray}
   \theta&=&\theta_{0}+\theta_{1}\epsilon+\theta_{2}\epsilon^{2}
          +\theta_{3}\epsilon^{3}+O(\epsilon^{4}),\\
   \mu&=&\mu_{0}+\mu_{1}\epsilon+\mu_{2}\epsilon^{2}
          +\mu_{3}\epsilon^{3}+O(\epsilon^{4}),
\end{eqnarray}
where the expansion parameter $\epsilon=\frac{\theta_{E}D_{OS}}{4D_{LS}}$ denotes the angle subtended by the gravitational radius normalized by the angular Einstein radius. It is easy to check that the coefficients $\theta_{0}\sim\theta_{2}$ and $\mu_{0}\sim\mu_{2}$ are independent of $q$, and their forms can be found in Ref.~\cite{Keeton}. Thus, the nonsingularity parameter $q$ only affects the position and magnification of images more than third-order in $\epsilon$. As a result, we come to the conclusion that the influence of the nonsingularity parameter can be ignored in the weak deflection limit.

\subsection{Relativistic images}
\label{UUU}

In this case, the spacetime always has a photon sphere. And a photon before reaching the observer can do many loops around the black hole, therefore the photon should pass very near the photon sphere. Adopting the method developed by Bozza, we define a variable \cite{Bozza}
\begin{eqnarray}
 z=\frac{f(r)-f(r_{0})}{1-f(r_{0})}.
\end{eqnarray}
For the photon at infinity $r=\infty$, one has $z=1$ for $f(\infty)=1$. And when $r=r_{0}$, one easily gets $z=0$. Then the total azimuthal angle (\ref{az}) can be rewritten as:
\begin{eqnarray}
 I(r_{0})=\int_{0}^{1}R(z,r_{0})K(z,r_{0})dz,\label{azimu}
\end{eqnarray}
where
\begin{eqnarray}
 R(z,r_{0})&=&\frac{2r_{0}(1-f(r_{0}))}{r^{2}f(r)f'(r)}
   =\frac{2r_{0}^{3}(r^{3}+q^{2})^{2}}
     {r^{3}(r_{0}^{3}+q^{2})(r^{3}-2q^{2})},\\
 K(z,r_{0})&=&\frac{r}{\sqrt{r^{2}f(r_{0})-f(r)r_{0}^{2}}}
   =\bigg(1+\frac{r_{0}^{2}}{r^{3}+q^{2}}
    -\frac{r_{0}^{2}}{r_{0}^{3}+q^{2}}-\frac{r_{0}^{2}}{r^{2}}\bigg)^{-1/2},
\end{eqnarray}
with $r=f^{-1}((1-f(r_{0}))z+f(r_{0}))$. Note that the function $R(z, r_{0})$ is regular for $z$ and $r_{0}$, while $K(z, r_{0})$ diverges at $z=0$. So, we split the integral (\ref{azimu}) into two parts:
\begin{eqnarray}
 I(r_{0})=I_{R}(r_{0})+I_{D}(r_{0}),
\end{eqnarray}
where the regular and divergent parts $I_{R}(r_{0})$, $I_{D}(r_{0})$ are, respectively, given by
\begin{eqnarray}
 I_{R}&=&\int_{0}^{1}g(z,r_{0})dx,\\
 I_{D}&=&\int_{0}^{1}R(0,r_{\text{ps}})K_{0}(z,r_{0})dz,
\end{eqnarray}
with $g(z,r_{0})=R(0,r_{0})K(z,r_{0})-R(0,r_{\text{ps}})K_{0}(z,r_{0})$.
In order to find the divergence of the
integrand, we do a Taylor expansion of the function inside the
square root of $K(z, r_{0})$, and obtain the function $K_{0}(z,r_{0})$:
\begin{eqnarray}
 K_{0}(z, r_{0})=\frac{1}{\sqrt{\chi(r_{0})z+\xi(r_{0})z^{2}+\mathcal{O}(z^{3})}},
\end{eqnarray}
where the coefficients $\chi(r_{0})$ and $\xi(r_{0})$ read
\begin{eqnarray}
 \chi(r_{0})&=&2-\frac{r_{0}^{2}}{r_{0}^{3}+q^{2}}
      -2\frac{r_{0}^{2}-3q^{2}}{r_{0}^{3}-2q^{2}},\\
 \xi(r_{0})&=&-1+3\frac{r_{0}^{8}-(6r_{0}-1)q^{2}r_{0}^{5}+3r_{0}^{3}q^{4}}
  {(r_{0}^{3}-2q^{2})^{3}}.
\end{eqnarray}
We find that, when $r_{0}$ approaches to $r_{\text{ps}}$, the coefficient $\chi(r_{0})$ vanishes, and the leading term of the divergence in $H_{0}$ is $z^{-1}$, which would lead to the logarithmic divergence of the integrand. Thus near $r_{\text{ps}}$, the deflection angle can be assumed in the form
\begin{eqnarray}
 \alpha(u)=-\bar{a}\log\big(\frac{u}{u_{\text{ps}}}-1\big)+\bar{b}
    +\mathcal{O}(u-u_{\text{ps}}). \label{Atheta}
\end{eqnarray}
Under this assumption, the minimum impact parameter $u_{\text{ps}}$, and the strong deflection limit coefficients $\bar{a}$ and $\bar{b}$ are
\begin{eqnarray}
 u_{\text{ps}}&=&\sqrt{\frac{r^{2}}{f(r)}}\bigg|_{r=r_{\text{ps}}}
   =\sqrt{\frac{r_{\text{ps}}^{3}+q^{2}}
    {r_{\text{ps}}^{2}(r_{\text{ps}}-1)+q^{2}/r_{\text{ps}}^{2}}},\label{ups39}\\
 \bar{a}&=&\frac{R(0,r_{\text{ps}})}{2\sqrt{\xi(r_{\text{ps}})}}
 =\sqrt{\frac{(r_{\text{ps}}^{3}+q^{2})(r_{\text{ps}}^{3}-2q^{2})}
          {(3-r_{\text{ps}})r_{\text{ps}}^{5}-11q^{2}r_{\text{ps}}^{3}+8q^{4}}},\\
 \bar{b}&=&-\pi+b_{R}+\bar{a}\log\kappa,\label{barb40}
\end{eqnarray}
%%%%%%%%%%%%%%%%%%%%%%%%%%%%%%%%%%%%%%%%%%%%%%%%%%%%%%%%%%%%%%%%%%%%%
\begin{figure}
\subfigure[]{\label{Ups}
\includegraphics[width=8cm]{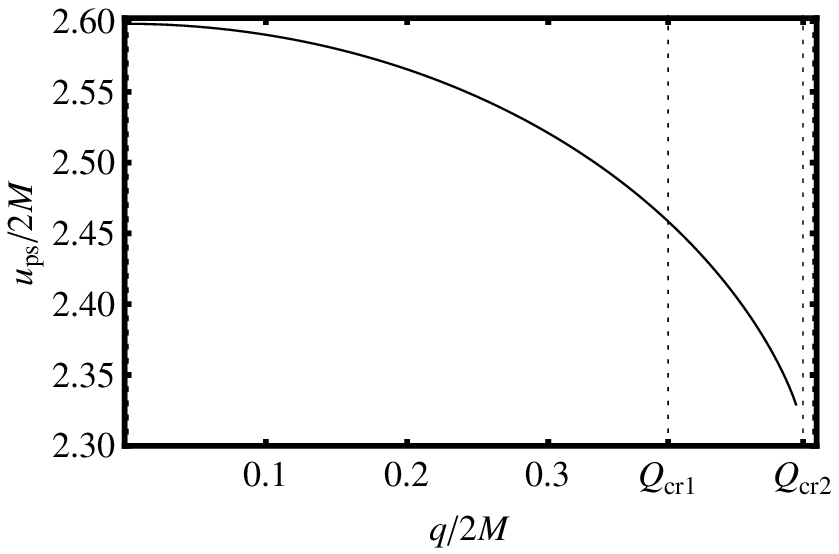}}
\subfigure[]{\label{Bbara}
\includegraphics[width=8cm]{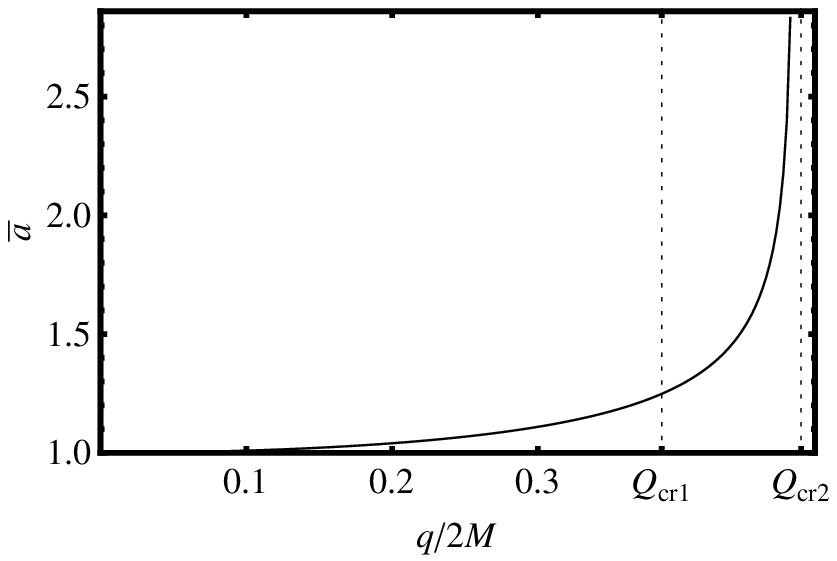}}\\
\centerline{\subfigure[]{\label{Barb}
\includegraphics[width=8cm]{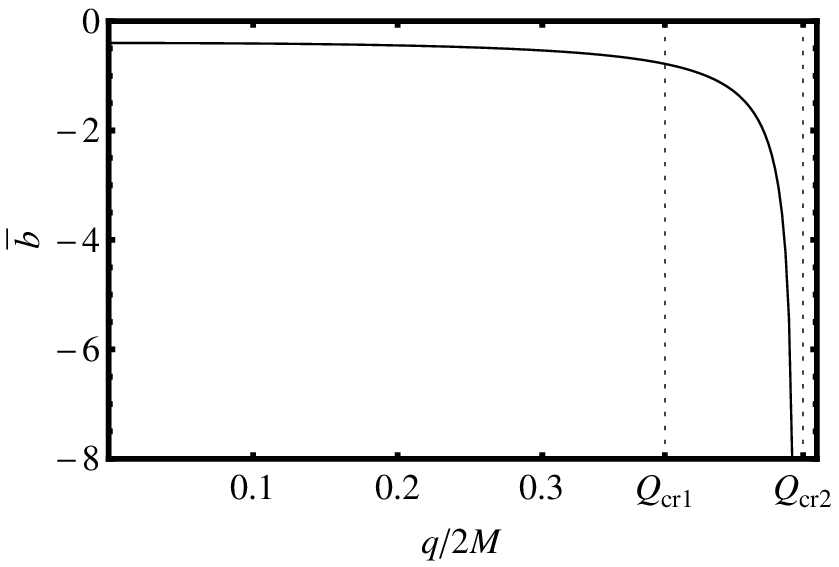}}}
\caption{Behaviors of the minimal impact parameter $u_{\text{ps}}$ and the strong deflection limit coefficients $\bar{a}$ and $\bar{b}$.} \label{coefficients1}
\end{figure}
%%%%%%%%%%%%%%%%%%%%%%%%%%%%%%%%%%%%%%%%%%%%%%%%%%%%%%%%%%%%%%%%%%%%%%%%%
%%%%%%%%%%%%%%%%%%%%%%%%%%%%%%%%%%%%%%%%%%%%%%%%%%%%%%%%%%%%%%%%%%%%%%%%%%%%%%
\begin{figure*}
\centerline{\subfigure[]{\label{Alpha1}
\includegraphics[width=8cm]{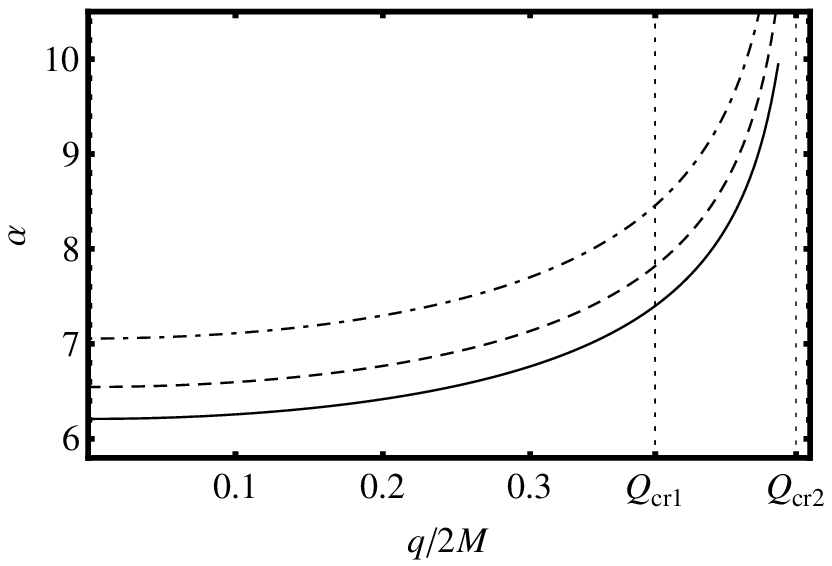}}
\subfigure[]{\label{Alpha2}
\includegraphics[width=8.2cm]{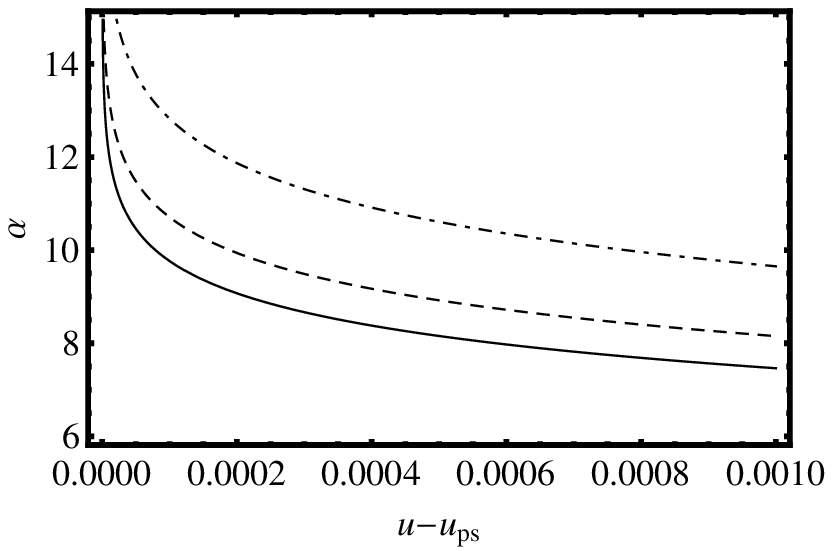}}}
\caption{(a) The deflection angle $\alpha$ as a function of $q/2M$ with $u=u_{\text{ps}}+0.0015$, $u_{\text{ps}}+0.0025$, $u_{\text{ps}}+0.0035$ from the bottom to up. (b) The deflection angle $\alpha$ as a function of $(u-u_{\text{ps}})$ for $q/2M$=0, 0.3, and 0.42 from the bottom to up.}
\label{PAlpha}
\end{figure*}
%%%%%%%%%%%%%%%%%%%%%%%%%%%%%%%%%%%%%%%%%%%%%%%%%%%%%%%%%%%%%%%%%%%%%%%%%%%%%%%%
where
\begin{eqnarray}
 \kappa=\frac{2\xi(r_{\text{ps}})}{f(r_{\text{ps}})}
     =\frac{2(r_{\text{ps}}^{3}+q^{2})^{2}((3-r_{\text{ps}})r_{\text{ps}}^{5}
 -11q^{2}r_{\text{ps}}^{3}+8q^{4})}
   {(r_{\text{ps}}^{3}-r_{\text{ps}}^{2}+q^{2})(r_{\text{ps}}^{3}-2q^{2})^{3}}.
\end{eqnarray}
For this nonsingular spacetime, the coefficient $b_{R}$ cannot be calculated analytically. In order to obtain it, we expand $I_{R}(r_{0})$ around $r_{\text{ps}}$,
\begin{eqnarray}
 I_{R}(r_{0})=\sum_{n=0}^{\infty}\frac{1}{n!}(r_{0}-r_{\text{ps}})^{n}
   \int_{0}^{1}\frac{\partial^{n}g}{\partial r_{0}^{n}}\bigg|_{r_{0}=r_{\text{ps}}}dz.
\end{eqnarray}
Therefore, ignoring the higher-order terms, we get
\begin{eqnarray}
 b_{R}=I_{R}(r_{\text{ps}})=\int_{0}^{1}g(z,r_{\text{ps}})dz.
\end{eqnarray}
With this equation, we can numerically calculate $b_{R}$. It is worth to point out that this result is accurate for the case $r_{0}\sim r_{\text{ps}}$, while invalid for $r_{0}\gg r_{\text{ps}}$. From these strong deflection limit coefficients, one easily see that there is a significant effect of the nonsingularity parameter $q$ on the strong gravitational lensing. When $q=0$, these parameters will reduce to the case of the Schwarzschild black hole \cite{Bozza,BozzaBozza}.

The behaviors of $u_{\text{ps}}$, $\bar{a}$, and $\bar{b}$ are presented in Fig. \ref{coefficients1}. The result shows that the minimum impact parameter has a similar behavior as the radius of the photon sphere. We can also find that the strong deflection limit coefficient $\bar{a}$ grows with $q/2M$, while $\bar{b}$ decreases. For a nonsingular black hole, both $\bar{a}$ and $\bar{b}$ have a finite value. Comparing with the result of nonsingular black hole, in the weakly nonsingular spacetime, it has small values of $u_{\text{ps}}$ and $\bar{b}$, while has large value of $\bar{a}$. In particular, $\bar{a}$ goes to positive infinity and $\bar{b}$ goes to negative infinity when the nonsingularity parameter $q/2M$ approaches $Q_{\text{cr2}}$, where the marginally nonsingular spacetime arrives. With the values of these coefficients, one can obtain the behavior of the deflection angle. For fixed impact parameter $u$, the deflection angle increases with $q/2M$ shown in Fig. \ref{Alpha1}. And in Fig. \ref{Alpha2}, it shows that $\alpha$ decreases with $(u-u_{\text{ps}})$ for fixed $q/2M$. However, the angle will be unboundedly large when $u\rightarrow u_{\text{ps}}$, which corresponds to the case that $r_{0}$ approaches to $r_{\text{ps}}$.

Here, we consider that the source, lens, and observer are highly aligned, i.e., $\varpi,\;\theta\ll 1$, the lens equation (\ref{lensgeometry}) is reduced to \cite{Bozza,BozzaBozza}
\begin{eqnarray}
 \varpi=\theta-\frac{D_{\text{LS}}}{D_{\text{OS}}}\Delta\alpha_{n},
 \label{varpi}
\end{eqnarray}
where $\Delta\alpha_{n}=\alpha-2n\pi$ and $n$ denotes the number of the loops that the photon done around the lens. Using the lens geometry, we can obtain the angular position and magnification of the images \cite{BozzaBozza,Bozza},
\begin{eqnarray}
 \theta_{n}&=&\theta_{n}^{0}+\frac{u_{\text{ps}}e_{n}}
         {\bar{a}}\frac{D_{\text{OS}}}{D_{\text{OL}}D_{\text{LS}}}
            (\varpi-\theta_{n}^{0}),\label{thetan}\\
 \mu_{n}&=&\frac{e_{n}(1+e_{n})}{\bar{a}\varpi}
         \frac{D_{\text{OS}}}{D_{\text{OL}}} \bigg(\frac{u_{\text{ps}}}{D_{\text{OL}}}\bigg)^{2}.\label{mun}
\end{eqnarray}
where
\begin{eqnarray}
 e_{n}&=&e^{\frac{\bar{b}-2n\pi}{\bar{a}}},\\
 \theta_{n}^{0}&=&\frac{u_{\text{ps}}}{D_{\text{OL}}}(1+e_{n}).
\end{eqnarray}
Since the magnification $\mu_{n}\sim e^{-n}$, the first image is the brightest one. We can also see that $\mu_{n}$ is proportional to the small quantity $\big(\frac{u_{\text{ps}}}{D_{\text{OL}}}\big)^{2}$, which leads to faint images. Thus, we have the conclusion that, for non-zero $\varpi$ the first image is the brightest one among these relativistic images and its brightness decreases quickly with the distance $D_{\text{OL}}$. On the other hand, it is worth to note that, when $\varpi=0$, the magnification (\ref{mun}) will be no longer valid.

\section{Numerical estimation of the observables for the supermassive Galactic black hole lensing}
\label{observational}

In this section, we first introduce three observables, and then estimate the numerical values for the observables of gravitational lensing in the strong field limit.

Consider the case that the outermost image with angular position $\theta_{1}$ is a single image and the others are packed together at $\theta_{\infty}=\frac{u_{\text{ps}}}{D_{\text{OL}}}$. Then we have the observables \cite{BozzaBozza},
\begin{eqnarray}
 s&=&\theta_{1}-\theta_{\infty}=\theta_{\infty}e^{\frac{\bar{b}-2\pi}{\bar{a}}}, \label{ss}\\
 \tilde{r}&=&\frac{\mu_{1}}{\sum_{n=2}^{\infty}\mu_{n}}=e^{2\pi/\bar{a}}. \label{rr},
\end{eqnarray}
where $s$ measures the angular separation between the first image and other ones, and $\tilde{r}$ denotes the flux of the first image and the sum of the others.

{\small
%%%%%%%%%%%%%%%%%%%%%%%%%%%%%%%%%%%%%%%%%%%%%%%%%%%%%%%%%%%%%%%%%%%%%%%%%%%%
\begin{table}[h]
\begin{center}
\begin{tabular}{|c|c|c|c|c|c|c|}
  \hline
     & $q/2M$=0 &  $q/2M$=0.2 & $q/2M$=0.3 & $q/2M$=0.39 & $q/2M$=0.42 & $q/2M$=0.45 \\
\hline
  $\bar{a}$  & 1.0000 & 1.0402 & 1.1094 & 1.2625 & 1.3771 & 1.6277 \\\hline
  $\bar{b}$& -0.4002 &  -0.4446 & -0.5369 & -0.8112 & -1.0782 & -1.8353 \\\hline
  $u_{\text{ps}}/R_{s}$& 2.5981  & 2.5659 & 2.5209 & 2.4537 & 2.4217 & 2.3806\\\hline
  $s$($\mu$\;arcsec) & 0.0211 & 0.0259 & 0.0350 & 0.0578 & 0.0750 & 0.1055 \\\hline
  $r_{m}$(magnitude)  & 6.8219 & 6.5583 & 6.1490 & 5.4036 & 4.9539 & 4.1911  \\\hline
  $\theta_{\infty}$($\mu$\;arcsec) & 16.8699  & 16.6610 & 16.3688 & 15.9321 & 15.7243 & 15.4580 \\\hline
\end{tabular}
\caption{Numerical estimation for main observables and the strong field limit coefficients for a Schwarzschild black hole, a nonsingular black hole, and a weakly nonsingular spacetime, which is supposed to describe the object at the center of our Galaxy. $r_{m}=2.5\log\tilde{r}$.}\label{tab1}
\end{center}
\end{table}}
%%%%%%%%%%%%%%%%%%%%%%%%%%%%%%%%%%%%%%%%%%%%%%%%%%%%%%%%%%%%%%%%%%%%%%%%%%%%%%

Next, we would like to estimate the numerical values for the observables of gravitational lensing in the strong field limit by supposing that, the gravitational field of the supermassive black hole at the center of our Milky Way can be described by the nonsingular metric (\ref{metric}), and the gravitational field near the light source and the observer is very weak so that it can be described by a flat metric. The mass of the supermassive black hole and the distance between the observer and the black hole are estimated to be $M=2.8\times 10^{6}M_{\odot}$ and $D_{\text{OL}}=8.5$ kpc with $M_{\odot}$ the mass of the sun. Under this assumption, the angular position can be calculated with the relation
\begin{eqnarray}
 \theta_{\infty}\approx 1.97116\times 10^{-5}u_{\text{ps}}
   \bigg(\frac{M}{M_{\odot}}\bigg)
     \bigg(\frac{1\text{kpc}}{D_{\text{OL}}}\bigg)\;\; \mu ~ \text{arcsec}.
\end{eqnarray}
From this equation, we get that the angular position $\theta_{\infty}$ depends on the parameter $u_{\text{ps}}$ and mass $M$ determined by the nature of the gravitational lens, and also on the observer-lens distance $D_{\text{OL}}$ determined by the lens geometry. In order to obtain a large value of $\theta_{\infty}$ for a same type lens, it should have large mass and small $D_{\text{OL}}$. For different values of the nonsingularity parameter $q$, the numerical values for the strong deflection limit coefficients and observables are listed in Table \ref{tab1}. It is clear that these results reduce to the Schwarzschild black hole spacetime when $q=0$. Moreover, we found that as $q$ increases, the angular position of the relativistic images $\theta_{\infty}$ and the relative magnitudes $r_{m}$ decrease, while the angular separation $s$ increases.

%%%%%%%%%%%%%%%%%%%%%%%%%%%%%%%%%%%%%%%%%%%%%%%%%%%%%%%%%%%%%%%%%%%%%
\begin{figure}
\subfigure[]{\label{Theta}
\includegraphics[width=8cm]{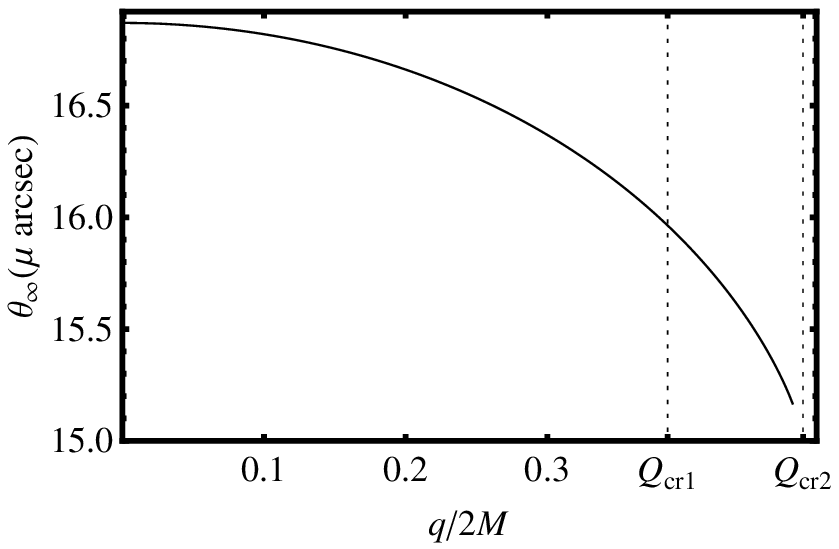}}
\subfigure[]{\label{Ss}
\includegraphics[width=8cm]{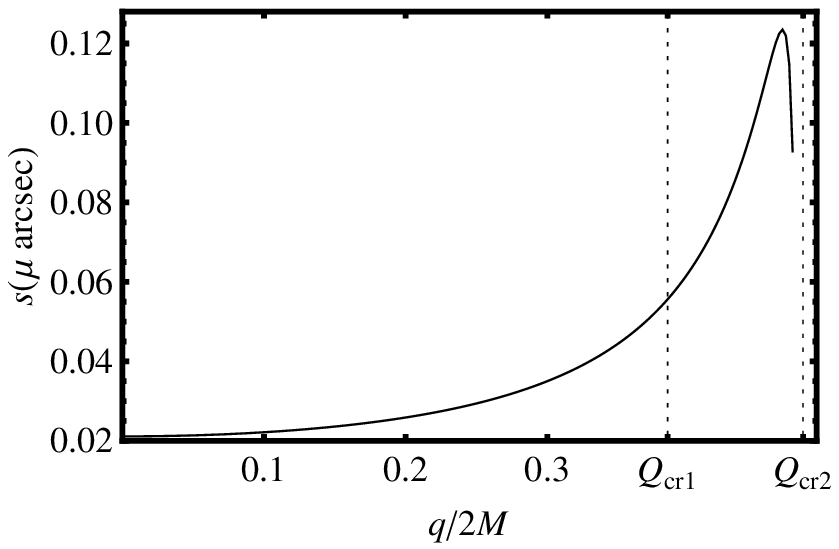}}\\
\centerline{\subfigure[]{\label{Rm}
\includegraphics[width=8cm]{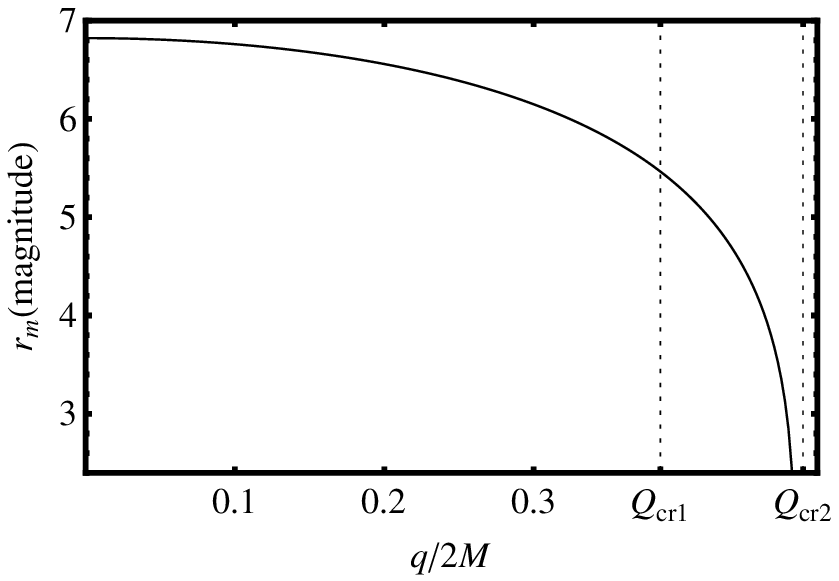}}}
\caption{Gravitational lensing by the Galactic center black hole. Variation of the values of the angular position $\theta_{\infty}$, angular separation $s$, and relative magnitudes $r_{m}$ with the parameter $q/2M$ in the nonsingular spacetime. $r_{m}=2.5\log\tilde{r}$.}\label{PObserver}
\end{figure}
%%%%%%%%%%%%%%%%%%%%%%%%%%%%%%%%%%%%%%%%%%%%%%%%%%%%%%%%%%%%%%%%%%%%%%%%%

The behaviors of the observables can also be found in Fig. \ref{PObserver}. We can find that, in the weakly nonsingular spacetime, the angular position $s$ of the relativistic images $\theta_{\infty}$ and the relative magnitudes $r_{m}$ decrease more quickly, and the angular separation $s$ increases more rapidly than the case of the nonsingular black hole. It is also clear that the angular position has a maximum value $s=0.1235\;\mu$arcsec at $q/2M=0.4659$. Comparing with a Schwarzschild black hole or a nonsingular black hole, a weakly nonsingular spacetime has a smaller angular position of the relativistic images and relative magnification of the outermost relativistic image with the other relativistic images. However, it has a larger angular separation for these relativistic images. From Fig. \ref{Theta}, we find that the numerical value for the angular position of the innermost relativistic images $\theta_{\infty}$ of a nonsingular spacetime is of about 15.3 $\sim$ 16.8$\mu arcsec$. In principle, such a resolution is reachable by very long baseline interferometry (VLBI) projects, and advanced radio interferometry between space and Earth (ARISE), which have the angular resolution of $10\sim 100$ $\mu arcsec$ in the near infrared \cite{Eckart,BozzaMancini}. Therefore we are hopeful to observe these relativistic images within a not so far future.

\section{Summary}
\label{Summary}

In this paper, we have shown that in the weak deflection limit, the influence of the nonsingularity parameter $q$ on the gravitational lensing is negligible. However, in the strong deflection limit, $q$ has a significant effect, which may offers a potentially powerful tool to probe the nonsingularity of spacetime.

First, we investigated the null geodesics of a nonsingular spacetime. According to the nature of the nonsingular spacetime, it is classified into several cases, such as the nonsingular black hole, the extremal nonsingular black hole, the weakly nonsingular spacetime, the marginally nonsingular spacetime, and the strongly nonsingular spacetime. We found that the photon sphere can exist not only in a nonsingular black hole background but also in a spacetime without a black hole. The result also shows that the photon is more easily captured by a nonsingular black hole rather than by a weakly nonsingular spacetime.

Second, based on the null geodesics, lensing in the weak and strong deflection limit was studied. For the first case, we found that the influence of the nonsingularity parameter $q$ on the positions and magnifications of the images is negligible. Thus, we cannot distinguish a nonsingular black hole from a Schwarzschild one. In the strong deflection limit, these strong deflection limit coefficients were also obtained. For a nonsingular black hole, these coefficients are always finite for any value of the nonsingularity parameter $q$. Comparing with the result of nonsingular black hole, in the weakly nonsingular spacetime, it has small values of $u_{\text{ps}}$ and $\bar{b}$, while has large value of $\bar{a}$. And when the nonsingular spacetime approaches to the marginally nonsingular one with $r_{\text{cir}}=r_{\text{ps}}$, we found that $\bar{a}$ goes to positive infinity and $\bar{b}$ goes to negative infinity. These results tell that the gravitational lensing by a weakly nonsingular spacetime is more obvious than a nonsingular black hole.

The model was also applied to the supermassive black hole hosted in the center of our Milky Way. It was shown that with the increase of the nonsingularity parameter $q/2M$, the angular position of the relativistic images $\theta_{\infty}$ and the relative magnitudes $r_{m}$ decrease, while the angular separation $s$ increases. We found that, in a weakly nonsingular spacetime, the angular position of the relativistic images $\theta_{\infty}$ and relative magnitudes $r_{m}$ decrease more quickly, and the angular separation $s$ increases more rapidly than a nonsingular black hole.

As pointed out in \cite{Virbhadra2009}, the relative magnitude $r_{m}$ obtained in this paper may have a large difference from its accurate value. However, the angular position $\theta_{\infty}$ and the angular separation $s$ are accurate enough. Thus, combining with the expected data from VLBI and ARISE, the strong gravitational lensing may offer a possible way to distinguish a weakly nonsingular spacetime from a nonsingular black hole. Furthermore, we are also allowed to probe the spacetime nonsingularity parameter $q$ by the astronomical instruments in the near future astronomical observations.

\section{Conflict of Interests}
The authors declare that there is no conflict of interests regarding the publication of this paper.

\section*{Acknowledgement}
The authors would like to thank the anonymous referees whose comments largely helped us in improving the original manuscript. S.-W. Wei wish to thank Dr. Changqing Liu for useful discussions and valuable comments. This work was supported by the National Natural Science Foundation of China (Grants No. 11205074, No. 11075065, and No. 11375075), and the Fundamental Research Funds for the Central Universities (Grants No. lzujbky-2013-21 and No. lzujbky-2013-18).

\end{document}